\documentclass[11pt,letterpaper]{article}
\usepackage[T1]{fontenc}
\usepackage[margin=1in,top=.9in,bottom=1in]{geometry}
\usepackage{amsmath,amsthm,amssymb}
\newtheorem{theorem}{Theorem}
\newtheorem{lemma}{Lemma}
\newtheorem*{matchingboundrestatement}{Lemma~\ref{lem:matching-cost}}
\usepackage{newtxtext,newtxmath}
\usepackage{microtype,graphicx,enumitem}
\usepackage[numbers,sort&compress]{natbib}
\usepackage[hidelinks]{hyperref}
\newcommand{\doi}[1]{doi:\space\href{https://doi.org/#1}{\nolinkurl{#1}}}
\hypersetup{
  pdftitle={A Better Approximation Algorithm for $k$-Delivery Traveling Salesman Problem},
  pdfsubject={Metric k-delivery TSP and preemptive versus nonpreemptive tours},
  pdfkeywords={k-delivery TSP, approximation algorithms, vehicle routing}
}
\setlist[enumerate]{label=(\arabic*),leftmargin=*,itemsep=.35em,topsep=.4em}

\title{A Better Approximation Algorithm for $k$-Delivery Traveling Salesman Problem}
\author{Jiajun He and Guochuan Zhang\footnote{zgc@zju.edu.cn}}
\date{(College of Computer Science, Zhejiang University)}
\makeatletter
\renewcommand{\maketitle}{%
  \begin{center}
    {\LARGE\bfseries \@title\par}
    \ifx\@author\@empty\else\vspace{.7em}{\normalsize \@author\par}\fi
    \ifx\@date\@empty\else\vspace{.4em}{\normalsize \@date\par}\fi
  \end{center}
  \vspace{.6em}
}
\makeatother

\begin{document}
\maketitle

\begin{abstract}
Given a metric graph in which each vertex either supplies or requires one item, where all items are identical, we consider the problem of transporting all items from pickup points to delivery points using a vehicle of capacity $k$. The objective is to minimize the total distance traveled by the vehicle. This problem generalizes the traveling salesman problem and is referred to as $k$-delivery TSP. The best known approximation algorithm achieves an approximation factor of 5 by \citet{CKR01}. Their method first constructs an approximate TSP tour, partitions it into segments according to the vehicle capacity, and matches the segments. The vehicle then travels between matched segments to complete the pickups and deliveries. This construction incurs round trips along matching edges and repeated travel within segments. It motivates us to reorganize the segments and matching edges to reduce the number of edge traversals, yielding a 4-approximation algorithm.

%There are two types of transportation: preemptive transportation allows items to be dropped at intermediate locations and picked up later, whereas nonpreemptive transportation does not allow such temporary storage. The former offers greater flexibility and may therefore yield a shorter route. Our algorithm constructs a nonpreemptive tour whose length is at most four times that of an optimal preemptive tour. Furthermore, we show that the length of an optimal nonpreemptive tour is at most three times that of an optimal preemptive tour.
\end{abstract}


\section{Introduction}\label{sec:introduction}

Vehicle routing problems (VRP) have been extensively studied in computer science and operations research. An important class consists of vehicle routing problems with pickups and deliveries (VRPPD), in which goods are transported from pickup points to delivery points without exceeding the vehicle capacity~\citep{PDH08}. These problems have preemptive and nonpreemptive versions. In a preemptive solution, items may be dropped at intermediate locations and picked up later to complete their delivery. In a nonpreemptive solution, an item is transported from its supply point to its demand point without being unloaded at intermediate locations. Allowing intermediate drops may change the nature of the problem and its algorithms~\citep{CKR01}. In this paper, we construct nonpreemptive tours and study the relationship between the optimal solutions of the two versions. The problem is formulated as the $k$-delivery traveling salesman problem ($k$-delivery TSP) below.

\paragraph{Problem statement.} Given an undirected graph $G=(V,E)$ with nonnegative edge costs $c(e)$ for $e\in E$, where the costs satisfy the triangle inequality, the set of vertices
\[
V=V^+\cup V^-
\]
is partitioned into a set $V^+$ of positive points and a set $V^-$ of negative points. Each positive point supplies one item, and each negative point requires one item. All items are identical, so any item can satisfy the demand of any negative point.

A vehicle of capacity $k$ picks up items at positive points and delivers them to negative points. We say that the vehicle \emph{serves} a point if it picks up an item at a positive point or delivers an item to a negative point. We may assume that $|V^+|=|V^-|$ and $k<|V^+|$. The objective is to transport all items from positive points to negative points without exceeding the vehicle capacity so that the cost of the tour is minimized. The starting point can be chosen arbitrarily, and the vehicle must start empty and return empty after completing all services.

\paragraph{Related work.}
Since TSP is a special case of $k$-delivery TSP, the latter is clearly NP-hard. The capacity constraint further requires us to reconcile the tour length with the order of pickups and deliveries. Although a short TSP tour connects all service points, serving them in the order of that tour need not be feasible: Too many consecutive pickups may exceed the vehicle capacity, whereas reaching a delivery point before collecting enough items may prevent the vehicle from serving it. Thus, we must both arrange a service order that satisfies the capacity constraint and bound the additional travel cost.

Previous approximation algorithms addressed this difficulty through tour partitioning and matching. \citet{CM99} constructed two approximate TSP tours, one containing all positive points and the other containing all negative points. They partitioned these tours into segments holding $k$ points each and used a matching between positive and negative segments to arrange the pickups and deliveries, obtaining a 9.5-approximation algorithm. The vehicle travels back and forth between the pickup tour and the delivery tour, combining the segments and matching edges into a feasible route. \citet{AB99} refined this approach to obtain a $(7-3/k)$-approximation algorithm.

Charikar, Khuller, and Raghavachari~\citep{CKR01} used a single TSP tour containing all positive and negative points, allowing pickups and deliveries to be combined on the same tour. They partitioned the tour into positive, negative, and zero segments according to cumulative supply and the capacity parameter, and computed a minimum-weight perfect matching between positive and negative segments. This approach exploits the proximity between supply and demand points. It achieves an approximation factor of 5 for even capacities and a bound of $31/6$ for odd capacities greater than one. They also showed that the optimal nonpreemptive tour length is at most four times the optimal preemptive tour length.

Prior work on special networks further exploited the structure of the underlying graph. Lim, Wang, and Xu~\citep{LWX05} gave a 2-approximation algorithm on trees. Wang, Lim, and Xu~\citep{WLX06} presented a polynomial-time optimal algorithm on paths and established the strong NP-hardness on trees. \citet{BH12} provided a linear-time optimal algorithm on paths and a $5/3$-approximation algorithm on trees of arbitrary height, while for trees of height two, the approximation ratio was reduced to $3/2-1/(2k)$. \citet{GZ19} showed a polynomial-time optimal algorithm on a circle and employed it to obtain a heuristic for general graphs. It first constructed an approximate TSP tour, then solved the resulting circle instance optimally. The theoretical approximation guarantee of this heuristic remains unclear.

These previous results show that the reference tour, its partition, and the way the segments are connected all affect the final service cost. In the tour-partitioning and matching framework of Charikar et al.~\citep{CKR01}, round trips along matching edges and repeated travel within segments incur additional costs. This motivates us to ask whether reorganizing the segments and matching edges, while retaining the partitioning and matching analysis, can reduce repeated travel and improve the approximation guarantee without violating the vehicle capacity. In this paper we present an affirmative answer.

\paragraph{Our results.}
For the $k$-delivery TSP, we present a better service route and show an approximation ratio at most 4. We also show that, given a preemptive tour, we can construct a nonpreemptive tour whose length is at most three times that of the given tour. Consequently, the optimal nonpreemptive tour length is at most three times the optimal preemptive tour length.

\section{Algorithmic Framework}\label{sec:outline}
\subsection{Overview of the Algorithm}\label{sec:overview}

Our algorithm starts from a TSP tour containing all positive and negative points. Building on the tour-partitioning and matching framework of \citet{CKR01}, we reorganize the segments and matching edges to reduce repeated travel. For the sake of convenience, we first consider even capacities $k$. The odd capacities are handled in Section~\ref{sec:analysis}. The algorithm consists of the following three steps.

\paragraph{Step 1. Partition the TSP tour into segments and match these segments.}
We fix a reference point and a traversal direction on the TSP tour, compute the cumulative net supply, and partition the tour according to the capacity parameter $k/2$. Namely, as soon as the net supply reaches one of the following three cases, a segment is created. A positive segment has a net supply of $k/2$ items, a negative segment has a net demand of $k/2$ items, and a zero segment has equal supply and demand. Clearly, the number of positive segments is equal to the number of negative segments as the supply and demand are balanced. We then compute a minimum-weight perfect matching between the positive and negative segments. Each matching edge joins a closest pair of points, one in each matched segment. If a partition contains only zero segments, we can serve all points by following the reference tour from a proper starting point. The next two steps handle partitions containing positive and negative segments.

\paragraph{Step 2. Combine the segments and matching edges into cycles.}
Step~1 pairs the positive and negative segments through the matching. Each path along the TSP tour connecting consecutive nonzero segments is called a \emph{gap}, which may contain zero segments. For simplicity, we regard each gap as a single edge. Note that there are an even number of such edges. We select alternate gaps around the TSP tour, yielding two complementary choices. For either choice, the positive and negative segments are connected by the matching edges and selected gap edges to form a collection of balanced cycles, each with zero net supply. We then construct a feasible route within each cycle, serving all its points while traversing every matching edge and selected gap edge of that cycle only once.

\paragraph{Step 3. Assemble the cycles into a complete tour.}
We contract each balanced cycle into a vertex and use the gaps not selected in Step~2 to connect these vertices. We choose a spanning tree of the resulting connected graph and recursively visit the cycles along the tree. Connections not included in the spanning tree are handled by additional out-and-back traversals. By coordinating the orientation of each cycle with the vehicle's load upon entry, we obtain a nonpreemptive tour that respects the capacity constraint. For the different partitions in Step~1 and the two complementary choices of gaps in Step~2, we construct candidate tours and return the shortest one.

This construction reduces repeated travel along matching edges, while introducing additional travel within segments and round trips between cycles. The next subsection gives the precise definitions of the partition and matching. Sections~\ref{sec:cycles} and~\ref{sec:assembly} present service within cycles and the global assembly, respectively, and establish their feasibility. Section~\ref{sec:analysis} accounts for all these costs, uses the complementary nature of the two constructions to bound the total length, and proves the final approximation guarantee.

\subsection{Tour Partitioning and Matching}\label{sec:framework}

In Step~1, we follow the tour-partitioning and matching framework of \citet{CKR01}. We fix a reference point and a traversal direction on the TSP tour containing all positive and negative points, and partition the tour into successive segments starting from that point.

When constructing a segment, we compute the cumulative net supply from its starting point. For the current edge $e$, let $\operatorname{excess}(e)$ denote the number of positive points minus the number of negative points encountered in the current segment before traversing $e$. Each time a positive point is encountered, the cumulative net supply increases by~$1$, while each time a negative point is encountered, it decreases by~$1$.

After at least one point has been encountered, the first time the cumulative net supply reaches $k/2$ or $-k/2$, or returns to~$0$, we cut the edge immediately after the current point to complete the segment. We then reset the cumulative net supply to~$0$ and start constructing the next segment from the next point. This process continues until the tour has been traversed. According to the cumulative net supply at the cut, the resulting segments are classified as positive, negative, or zero. A positive segment has a net supply of $k/2$ items, a negative segment has a net demand of $k/2$ items, and a zero segment has zero net supply.

Since each segment ends at the first occurrence of one of these three values, every nonempty proper prefix of a positive segment has net supply strictly between $0$ and $k/2$, and every nonempty proper prefix of a negative segment has net supply strictly between $-k/2$ and~$0$. For a zero segment, the net supplies of all its nonempty proper prefixes lie either entirely in $(0,k/2)$ or entirely in $(-k/2,0)$.

Recall that there are equally many positive and negative segments. We construct a weighted bipartite graph whose two sides correspond to positive and negative segments. The weight of an edge between a positive segment and a negative segment is the minimum distance between a point in one segment and a point in the other. We then compute a minimum-weight perfect matching in this bipartite graph.

For each pair of matched segments, we select two points attaining this minimum distance and call the edge joining them a matching edge. The next section explains how to construct cycles with zero net supply from these segments and their matching, and how to arrange a feasible service order within each cycle.

\section{Constructing and Servicing Cycles}\label{sec:cycles}
\subsection{Constructing Balanced Cycles}\label{sec:cycle-construction}

For a partition containing nonzero segments and its matching from Step~1, our goal is to reorganize these segments, the matching edges, and edges of the TSP tour into balanced cycles with zero net supply, so that each matching edge appears exactly once in its cycle.

Since there are equally many positive and negative segments, the total number of nonzero segments is even. Let $S_1,S_2,\ldots,S_{2r}$ denote these segments in their cyclic order on the TSP tour. The path along the TSP tour between two consecutive nonzero segments is called a gap and may contain zero segments. For simplicity, we regard each gap as a single edge.

\begin{figure}[htbp]
  \centering
  \includegraphics[width=.95\linewidth]{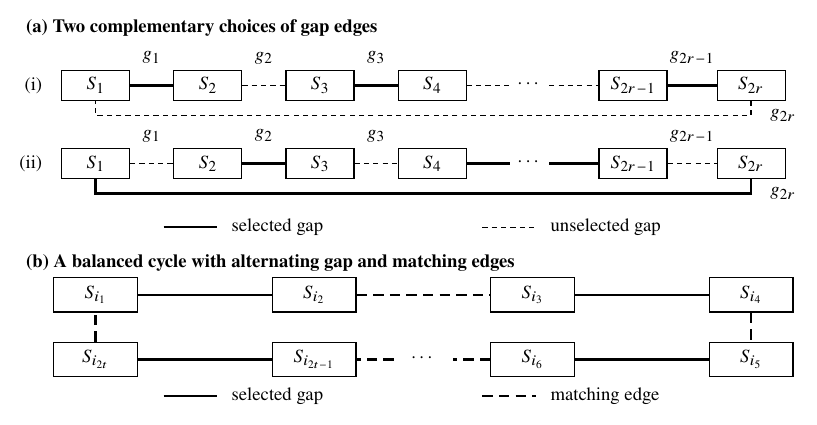}
  \caption{(a) Two complementary choices of gaps. (b) A balanced cycle with alternating matching and gap edges. Boxes represent nonzero segments.}
  \label{fig:gap-choices}
\end{figure}

As illustrated in Figure~\ref{fig:gap-choices}(a), we consider two complementary pairings of consecutive segments. One pairs $S_1$ with $S_2$, $S_3$ with $S_4$, $\ldots$, and $S_{2r-1}$ with $S_{2r}$, while the other pairs $S_2$ with $S_3$, $S_4$ with $S_5$, $\ldots$, and $S_{2r}$ with $S_1$. For each pairing, we retain the gaps connecting the paired segments. The remaining gaps are not used to construct the balanced cycles and are left for the subsequent assembly. Thus, each nonzero segment is incident to exactly one retained gap, and each gap is retained in exactly one of the two choices. For a given partition (that is, a fixed reference point and traversal direction), the two complementary choices of gaps use the same matching edges and matching endpoints.

We regard each nonzero segment as a vertex, with the matching edges and retained gaps forming the edges between these vertices. Each vertex is incident to exactly one matching edge and one gap edge. Hence, the resulting graph consists of cycles in which matching edges and gap edges alternate, as illustrated in Figure~\ref{fig:gap-choices}(b). Each matching edge and each retained gap edge appears exactly once in one of these cycles.

The matching edges in each cycle pair its positive and negative segments one to one, so the cycle contains equally many positive and negative segments. Each positive segment has net supply $k/2$, each negative segment has net demand $k/2$, and the zero segments located within gaps have zero net supply. Therefore, the total net supply of each cycle is zero. We call these cycles \emph{balanced}.

\subsection{Serving a Cycle}\label{sec:cycle-service}

\begin{figure}[htbp]
  \centering
  \includegraphics[width=.95\linewidth]{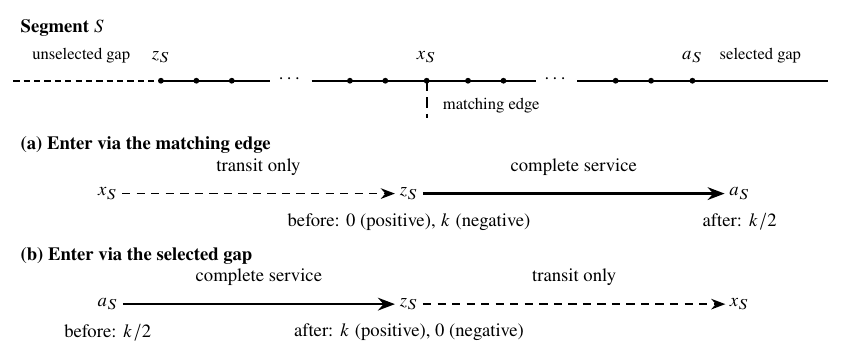}
  \caption{Service order and endpoint loads. Solid arrows indicate complete service; dashed arrows indicate travel without service.}
  \label{fig:segment-service}
\end{figure}

For each balanced cycle, choose an arbitrary traversal direction. For a nonzero segment $S$ on the cycle, let $a_S$ denote its endpoint incident to the retained gap, $z_S$ its endpoint incident to the unselected gap, and $x_S$ the endpoint of its matching edge, as shown in the top diagram of Figure~\ref{fig:segment-service}. We arrange service according to the edge through which the vehicle enters the segment.

If the vehicle enters through the matching edge, it first travels along $S$ from $x_S$ to $z_S$ without performing any pickups or deliveries. It then traverses the entire segment from $z_S$ to $a_S$, performing all pickups and deliveries, and leaves through the retained gap, as shown in Figure~\ref{fig:segment-service}(a).

If the vehicle enters through the retained gap, it first traverses the entire segment from $a_S$ to $z_S$, performing all pickups and deliveries. It then travels along $S$ from $z_S$ to $x_S$ without further service and leaves through the matching edge, as shown in Figure~\ref{fig:segment-service}(b). Zero segments within retained gaps are served as the vehicle traverses them. Points at the ends of a gap are served as part of the adjacent nonzero segments.

Following this order, a complete traversal of the balanced cycle serves all its points, with each matching edge and each retained gap edge traversed exactly once.

\begin{lemma}[Cycle service and endpoint loads]\label{lem:cycle-loads}
The above cyclic service order admits a load profile that respects the capacity constraint. At $a_S$, the vehicle has load $k/2$ when entering or leaving the segment through the retained gap. At $z_S$, the load is $0$ or $k$ immediately before completing service if the vehicle enters through the matching edge, or immediately after completing service if it enters through the gap.
\end{lemma}

\begin{proof}
For a positive segment, the loads before and after completing service are $0\to k/2$ and $k/2\to k$ in the two cases above, respectively. For a negative segment, they are $k\to k/2$ and $k/2\to0$. By the prefix net supply properties in Section~\ref{sec:framework}, the load remains in $[0,k]$ throughout service, while travel without service does not change the load.

Each matching edge joins segments of opposite signs, so the load when leaving one segment along the matching edge is exactly the load required to enter the other. Each zero segment within a retained gap can be served starting with load $k/2$, respecting the capacity constraint and restoring the load to $k/2$. Thus, the load assignments agree across all segments and gaps around the cycle, and a complete traversal restores the entering load.
\end{proof}

Furthermore, reverse the order of all travel and service operations around the cycle, and replace the load $q$ at each corresponding instant by $k-q$. A pickup still increases the load by~$1$, and a delivery still decreases it by~$1$, so the reversed service order is also feasible and has the same cost. The load at $a_S$ when entering or leaving through the retained gap remains $k/2$, while the empty and full states at the corresponding instants at $z_S$ are exchanged.

Consequently, whether the vehicle reaches any specified $z_S$ on a balanced cycle with load $0$ or $k$, we can choose a service direction that completes all services on the cycle while respecting the capacity constraint throughout. In the subsequent assembly, service of the first balanced cycle can start empty at any $z_S$. The vehicle also enters every other balanced cycle at a $z_S$ and chooses its service direction according to the entering load.

\section{Joining the Cycles into a Tour}\label{sec:assembly}

In Step~3 of the algorithm, we use the unselected gap edges to join the balanced cycles constructed in the previous step into a tour that keeps the capacity constraint valid.

\subsection{The Connection Graph and a Spanning Tree}\label{sec:connection-tree}

Contract each balanced cycle into a vertex and use the unselected gaps as edges between these vertices. The retained and unselected gaps together connect all consecutive nonzero segments along the original TSP tour, and every retained gap lies within a balanced cycle. Hence, the resulting connection graph is connected. After contraction, distinct unselected gaps may join the same pair of cycles, or a gap may join two positions on the same cycle. We keep all these edges.

\begin{figure}[htbp]
  \centering
  \includegraphics[width=.95\linewidth]{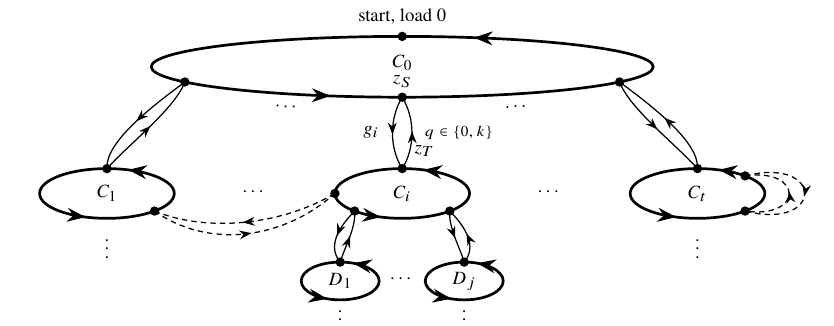}
  \caption{Tree assembly at the $z_S$ positions (dots). Thick arrows show cyclic service order; paired thin arrows show gap round trips. Non-tree gaps are dashed.}
  \label{fig:tree-assembly}
\end{figure}

Choose a spanning tree of the connection graph and an arbitrary balanced cycle $C_0$ as its root. Choose any $z_S$ on $C_0$ as the starting point and, by Section~\ref{sec:cycle-service}, orient its service so that the vehicle starts empty. As illustrated in Figure~\ref{fig:tree-assembly}, the spanning tree determines the parent--child relationships between cycles. Gaps outside the spanning tree are handled by additional out-and-back traversals.

\subsection{Recursive Traversal and Feasibility}\label{sec:recursive-traversal}

Starting from the root cycle, the vehicle follows the service order in Section~\ref{sec:cycle-service}. At $z_S$, immediately before beginning complete service of the segment or immediately after completing it, if the incident tree edge leads to an unvisited child cycle, the vehicle suspends service of the current cycle and travels along the corresponding gap to the child. According to its load on arrival, either $0$ or $k$, it chooses the child's service direction and recursively serves that cycle and all its descendants. It then returns along the same gap and resumes service of the parent cycle. Children are visited in the order in which their attachment positions occur in the parent's service order.

For each gap outside the spanning tree, choose one endpoint to handle it. At the corresponding service instant at that endpoint, the vehicle traverses the gap out and back before resuming service of its current cycle. This excursion serves only the zero segments within the gap; it does not initiate service of the cycle at the other endpoint. Gaps joining two positions on the same cycle are handled in the same way. Each such gap is assigned exactly one out-and-back traversal.

For every unselected gap, whether or not it belongs to the spanning tree, we serve its zero segments as follows. If the vehicle enters the gap with load $0$, serve each zero segment in the direction with nonnegative prefix net supply; if it enters with load $k$, use the direction with nonpositive prefix net supply. Since the outward and return traversals have opposite directions, one of them permits this service; on the other, the vehicle travels through the segment without performing any pickups or deliveries. This choice is made separately for each zero segment.

By the prefix net supply properties in Section~\ref{sec:framework}, service in the chosen direction respects the capacity constraint. Each zero segment has zero net supply, so its service restores the entering load, while the traversal without service leaves the load unchanged. Zero segments within balanced cycles are served with their cycles as described in Section~\ref{sec:cycle-service}.

\begin{lemma}[Feasibility of the global assembly]\label{lem:assembly-feasibility}
The above assembly produces a nonpreemptive tour that respects the capacity constraint. The vehicle starts empty and returns empty, and every service point is served exactly once. Each matching edge and each retained gap is traversed once, and each unselected gap is traversed twice.
\end{lemma}

\begin{proof}
We prove the following invariant by induction on subtrees of the spanning tree. If the vehicle enters a cycle with load $q\in\{0,k\}$, then, after serving that cycle and all its descendants and completing the associated gap excursions, it returns to the same entry position with load $q$.

For a leaf cycle, both the cycle service in Section~\ref{sec:cycle-service} and the non-tree gap excursions assigned to it restore the entering load. For an internal cycle, the induction hypothesis ensures that each visit to a child subtree restores the load with which the vehicle entered the child cycle. After returning along the connecting gap, the vehicle therefore has the same load at the parent position as when it left, and can resume the suspended service. All operations respect the capacity constraint. Since the root cycle is entered with load $0$, the resulting tour starts empty and returns empty.

Each nonzero segment is completely served exactly once within its own cycle. Zero segments within retained gaps are served with their cycles, and those within unselected gaps are served on exactly one of the two traversals. Thus, every point is served exactly once. %All load changes arise from pickups at supply points or deliveries at demand points, without temporary storage, so the tour is nonpreemptive. 
Service of each nonroot cycle is initiated exactly once through its unique parent edge, and every unselected gap is assigned exactly one out-and-back traversal. This completes the proof.
\end{proof}

\section{Analysis}\label{sec:analysis}
%\subsection{Cost of the Assembled Tour}\label{sec:assembly-cost}

The complementary nature of the two choices of gap edges gives the following bound.

\begin{lemma}[Cost of the assembled tour]\label{lem:assembly-cost}
For a partition containing nonzero segments, let $T$ be the cost of the reference TSP tour and $M$ the cost of the corresponding matching. Of the two tours constructed in Section~\ref{sec:cycles} and~\ref{sec:assembly} using the complementary choices of gaps, at least one has cost at most
\[
\frac32T+M.
\]
\end{lemma}

\begin{proof}
The two constructions use the same matching edges and matching endpoints. Let $W_0$ and $W_1$ denote the costs of the resulting tours. For each tour, first count one complete service traversal of every nonzero segment and one traversal of every gap. Together, these parts form the original TSP tour and have cost $T$, including travel through the zero segments within gaps. Each matching edge is traversed once, at a total cost of $M$. There are two additional costs: travel without service along each nonzero segment between $x_S$ and $z_S$, and the second traversal of each unselected gap.

Switching between the two complementary choices exchanges $a_S$ and $z_S$ for every nonzero segment, while $x_S$ remains fixed. Thus, the additional travel within that segment connects $x_S$ to one endpoint in one tour and to the other endpoint in the other tour. The two subpath lengths sum to the length of the segment. Moreover, each gap is unselected in exactly one of the two choices, so it contributes one additional traversal across the two tours. Since the nonzero segments and gaps together form the reference TSP tour, the additional costs of the two tours sum to $T$. Hence,
\begin{equation}\label{eq:complementary-tour-cost}
W_0+W_1=2(T+M)+T=3T+2M.
\end{equation}
Taking the cheaper tour gives
\[
\min\{W_0,W_1\}\leq\frac32T+M.
\]
\end{proof}

%\subsection{The 4-Approximation Guarantee}\label{sec:approximation-guarantee}

We combine the matching cost with the assembly cost to prove the 4-approximation guarantee. For $k\geq2$, let $h=\lfloor k/2\rfloor$ and run the construction with effective capacity $2h$. For even capacities, the construction is unchanged. For odd capacities, the partitioning threshold becomes $h$, and the half-full and full loads in Sections~\ref{sec:cycles} and~\ref{sec:assembly} become $h$ and $2h$, respectively. Since $2h\leq k$, the resulting tour respects the original capacity constraint.

Let $T$ be the cost of the reference TSP tour. We fix one traversal direction and enumerate the distinct partitions obtained by taking each service point as the reference point. If some partition contains only zero segments, we can serve all points by following the reference tour from a suitable starting point, at cost $T$. Otherwise, we use the following matching cost bound, where $B$ denotes the cost of a minimum-weight perfect matching between the original positive and negative points.

\begin{lemma}[Matching cost bound]\label{lem:matching-cost}
If every enumerated partition contains nonzero segments, then at least one of these partitions has a minimum-weight perfect matching between its positive and negative segments whose cost $M$ satisfies
\[
M\leq\frac Bh+\frac{h-1}{2h}T.
\]
\end{lemma}

This lemma follows the local matching and averaging argument of \citet{CKR01}. Appendix~\ref{app:proofs} gives a detailed proof.

The algorithm returns the cheapest of all candidate tours. Let $W$ denote its cost. By Lemma~\ref{lem:assembly-cost},
\begin{equation}\label{eq:reference-tour-bound}
W\leq\frac32T+M
\leq\left(2-\frac1{2h}\right)T+\frac Bh.
\end{equation}
If some partition contains only zero segments, the reference tour can be served at cost $T$, which also satisfies the final bound.

\begin{theorem}[4-approximation guarantee]\label{thm:four-approximation}
For $k$-delivery TSP, the algorithm constructs in polynomial time a nonpreemptive tour whose cost is at most four times that of an optimal preemptive tour.
\end{theorem}

\begin{proof}
For any preemptive tour of cost $C$, the optimal TSP tour cost $T^*$ and the matching cost $B$ satisfy
\[
T^*\leq C,\qquad 2B\leq kC.
\]
The first inequality follows from the triangle inequality. The second uses both the actual load and the remaining vehicle capacity; its proof is given in Appendix~\ref{app:lower-bounds}. Constructing the reference tour with the algorithm of Christofides~\citep{Christofides76} gives $T\leq3C/2$. Thus, for $k\geq2$, equation~\eqref{eq:reference-tour-bound} yields
\[
\begin{aligned}
W
&\leq\left[\frac32\left(2-\frac1{2h}\right)+\frac{k}{2h}\right]C\\
&=\left(3+\frac{2k-3}{4h}\right)C
\leq4C,
\end{aligned}
\]
where the last inequality uses $k\leq2h+1$.

For $k=1$, shortcut the reference tour to a tour visiting only the positive points. At each positive point, pick up its item, traverse the corresponding edge of the minimum-weight perfect matching between the original positive and negative points to deliver it, and return to that positive point before continuing along the tour. This construction starts and ends empty and respects the capacity-one constraint throughout. Its cost is at most
\[
T+2B\leq\frac32C+C\leq4C.
\]
Since these bounds hold for every preemptive tour, the resulting nonpreemptive tour costs at most four times an optimal preemptive tour.
\end{proof}

It remains to analyze the running time. Let $N=|V|$. For $k\geq2$, let $h=\lfloor k/2\rfloor$. After fixing a traversal direction of the reference tour, the algorithm considers at most $h$ distinct partitions, processing each partition only once; see Appendix~\ref{app:matching-proof}. If a partition contains $r$ positive segments and $r$ negative segments, then each nonzero segment contains at least $h$ service points, so
\[
r\leq\frac{N}{2h}.
\]

For each partition, scanning the reference tour to form the segments and gaps takes $O(N)$ time. By examining pairs of points, we can compute all distances between positive and negative segments and record the corresponding endpoints in $O(N^2)$ time. A minimum-weight perfect matching between the segments can be computed in $O(r^3)$ time~\citep{Gabow90}. Once the matching is fixed, constructing the balanced cycles, computing a spanning tree, and recursively assembling the tour take $O(N)$ time for each of the two complementary choices of gaps. We output travel and service operations directly during the recursive traversal. Since each segment and gap is traversed only a constant number of times, the representation of each candidate tour has length $O(N)$.

Thus, processing all partitions takes
\[
O\!\left(hN^2+h\left(\frac{N}{2h}\right)^3\right)
=O\!\left(hN^2+\frac{N^3}{h^2}\right)
=O(N^3),
\]
where we use $h\geq1$ and $h=O(N)$. Including the $O(N^3)$ time required to construct the reference TSP tour, the overall running time is $O(N^3)$. For $k=1$, a minimum-weight perfect matching between the original positive and negative points can also be computed in $O(N^3)$ time, so the same bound holds.

Processing candidate tours one at a time, we need only store the distance matrix, the current matching instance, and the current and best tours, requiring $O(N^2)$ space. Therefore, our algorithm has polynomial time and space complexity. Further implementation details and the bit complexity for rational inputs are discussed in Appendix~\ref{app:implementation}.

\paragraph{The Preemption Gap}\label{sec:preemption-gap}

The cost bound from the preceding analysis also gives a threefold conversion guarantee from preemptive tours to nonpreemptive tours.

\begin{theorem}[Threefold conversion guarantee]\label{thm:preemption-gap}
Given a preemptive tour of cost $C$, one can construct in polynomial time a nonpreemptive tour of cost at most $3C$. Consequently, the cost of an optimal nonpreemptive tour is at most three times that of an optimal preemptive tour.
\end{theorem}

\begin{proof}
Shortcut the given preemptive tour in the order in which the service points first appear to obtain a reference TSP tour containing all service points. By the triangle inequality, its cost satisfies $T\leq C$.

For $k\geq2$, let $h=\lfloor k/2\rfloor$ and run the construction with effective capacity $2h$. By equation~\eqref{eq:reference-tour-bound} in Section~\ref{sec:analysis} and the lower bound $2B\leq kC$ from Lemma~\ref{lem:preemptive-lower-bounds}, the cost $W$ of the resulting nonpreemptive tour satisfies
\[
\begin{aligned}
W
&\leq\left(2-\frac1{2h}\right)T+\frac Bh\\
&\leq\left(2+\frac{k-1}{2h}\right)C
\leq3C,
\end{aligned}
\]
where the last inequality uses $k-1\leq2h$.

For $k=1$, the capacity-one construction in Section~\ref{sec:analysis} has cost at most
\[
T+2B\leq2C\leq3C.\qedhere
\]
\end{proof}

\section{Conclusions}

This paper revisited the 5-approximation algorithm \cite{CKR01} for the $k$-delivery TSP and found a better way to make a feasible tour, which results in an approximation bound of 4. It is interesting to find an alternative approach to beat the current ratio and figure out lower bounds beyond the approximation hardness for TSP.

%\clearpage
\section*{AI Usage Statement}
The authors provide the algorithm structure, key proof steps, and the initial draft. They used OpenAI's GPT-6 Astra help refine the proof details, polish the presentation, and improve LaTeX formatting. The authors independently verified all mathematical claims, proofs, and references and take full responsibility for the manuscript.

\bibliographystyle{plainnat}
\bibliography{references}

\clearpage
\appendix
\section{Omitted Proofs}\label{app:proofs}
\subsection{Proof of Lemma~\ref{lem:matching-cost}}\label{app:matching-proof}

For $k\geq2$, let $h=\lfloor k/2\rfloor$, let $T$ be the cost of the reference TSP tour, and let $B$ be the cost of a minimum-weight perfect matching between the original positive and negative points. We restate the bound to be proved.

\begin{matchingboundrestatement}[Matching cost bound, restated]
If every enumerated partition contains nonzero segments, then at least one of these partitions has a minimum-weight perfect matching between its positive and negative segments whose cost $M$ satisfies
\[
M\leq\frac Bh+\frac{h-1}{2h}T.
\]
\end{matchingboundrestatement}

We first establish the partition correspondence, then use auxiliary matchings within segments to bound the matching cost for a fixed partition, and finally average over the partitions.

\paragraph{Partition correspondence.}
Fix a starting point and a traversal direction on the reference TSP tour. For each edge $e$, let $F(e)$ be the number of positive points minus the number of negative points encountered from this starting point before traversing $e$. This count is maintained over the entire tour and is not reset at the end of a segment. For $i\in\{0,\ldots,h-1\}$, consider the partition obtained by cutting all edges satisfying
\[
F(e)\equiv i\pmod h,
\]
initially considering only residue classes that occur on the tour.

These partitions are precisely the distinct partitions enumerated by our algorithm. Indeed, if the edge preceding the reference point of a partition has $F$-value $c$, then the cumulative net supply within the first segment prior to any edge $e$ is $F(e)-c$. Since the cumulative value changes by only $1$ or $-1$ at each point, the first subsequent value congruent to $c$ modulo $h$ occurs when $F$ first reaches $c-h$, $c$, or $c+h$. This agrees with the rule in Section~\ref{sec:framework}, which cuts at the first occurrence of $-h$, $0$, or $h$. After each cut, the new starting cumulative value remains in the same residue class, so repeating the rule segment by segment cuts exactly all edges in that class. The edge preceding the reference point also belongs to this class, and the last segment ends at that edge. Conversely, choosing the service point immediately after any edge in a class as the reference point yields the partition for that class. Thus, for a fixed direction, there are at most $h$ distinct partitions; different reference points corresponding to the same class only cyclically shift the order of the segments.

If a partition contains only zero segments, suppose that the vehicle starts at the beginning of any segment carrying $h$ items and performs all pickups and deliveries in order along the reference tour. By the prefix property in Section~\ref{sec:framework}, each zero segment begins and ends with load $h$, and the load throughout its service lies in $[0,2h]$. Now shift the starting point to a position where the load is minimized and start empty. The new load equals the original load minus this minimum value, so it remains in $[0,2h]$ and returns to zero after one complete traversal. Since $2h\leq k$, this yields a feasible nonpreemptive tour of cost $T$.

If some residue class modulo $h$ does not occur, then the difference between the maximum and minimum values of $F$ is less than $h$; otherwise, the unit changes of $F$ would pass through consecutive integer values representing all residue classes. In this case, no segment can have net supply $h$ or $-h$, so every partition that arises contains only zero segments and is covered by the preceding case.

Therefore, under the hypothesis of Lemma~\ref{lem:matching-cost}, all $h$ partitions exist, and each contains nonzero segments. We next fix one of these partitions and analyze its matching cost, then average over the $h$ partitions, counting each exactly once.

\paragraph{Matching cost for a fixed partition.}
For a fixed partition, we relate the matching cost between segments to the matching cost between the original positive and negative points by constructing an auxiliary matching within the segments.

\begin{lemma}[Matching cost for a fixed partition]\label{lem:fixed-partition-matching}
Fix the $i$th partition, which contains nonzero segments, and let $M_i$ be the cost of a minimum-weight perfect matching between its positive and negative segments. There is an auxiliary matching within the segments that matches all points in zero segments and leaves exactly $h$ unmatched positive points in each positive segment and $h$ unmatched negative points in each negative segment. If this auxiliary matching has cost $A_i$, then
\[
M_i\leq\frac{B+A_i}{h}.
\]
\end{lemma}

\begin{proof}
We first construct the auxiliary matching within the segments and then establish the cost bound.

For a zero segment, choose the traversal direction in which every prefix has nonnegative net supply; if the nonempty proper prefixes have negative net supply in the original direction, reverse the direction. Starting with an empty stack, push each positive point onto the stack, and match each negative point to the positive point at the top of the stack before popping that positive point. The nonnegative prefix net supply ensures that no negative point is encountered when the stack is empty, and the zero net supply of the segment ensures that the stack is empty at the end. Thus all points are matched. For each stack position, record in order the service points that occupy or empty that position. This gives an alternating sequence of positive and negative points, beginning with a positive point and ending with a negative point. The matching pairs each positive point in this sequence with the negative point immediately following it. By the triangle inequality, the cost of each matching edge is at most the length of the corresponding subpath within the segment. These subpaths are disjoint within a sequence, so the matching cost for that sequence is at most the length of the subpath from its first point to its last point.

For a positive segment, use the same push and pop rules in the fixed direction. By the prefix property, the stack size always lies in $[0,h]$ and is exactly $h$ at the end of the traversal. Thus the sequence of service points for each stack position alternates between positive and negative points, beginning and ending with a positive point. For each such sequence, consider two ways to pair adjacent points: pair consecutive points starting with the first point, leaving the last positive point unmatched, or start with the second point, leaving the first positive point unmatched. The sum of the costs of these two matchings equals the length of the path joining consecutive points in the sequence. By the triangle inequality, this length is at most that of the subpath of the original segment from the first point of the sequence to its last point. Choose the cheaper matching, whose cost is at most half the length of this subpath. One positive point remains unmatched at each stack position, leaving exactly $h$ unmatched positive points in the segment. If a sequence consists of a single point, no matching edge is added and its cost is zero.

The construction for a negative segment is symmetric: push negative points and pop at positive points. Each stack position gives an alternating sequence beginning and ending with a negative point. Again choose the cheaper of the two adjacent pairings. This leaves exactly $h$ unmatched negative points in each negative segment, and the matching cost for each sequence is at most half the length of the corresponding subpath within the segment. Each service point belongs to one segment and appears in the sequence of exactly one stack position. Combining these matchings therefore gives the required auxiliary matching; denote its cost by $A_i$.

Superimpose this auxiliary matching and a minimum-weight perfect matching of cost $B$ between the original positive and negative points, retaining two copies if an edge belongs to both matchings. Points matched by the auxiliary matching have degree two in the resulting multigraph, whereas points left unmatched by the auxiliary matching have degree one. Hence every connected component is a path or a cycle whose edges alternate between the two matchings. The first and last edges of every path belong to the original point matching, so the path has an odd number of edges and its endpoints have opposite signs. These endpoints are unmatched points left in a positive segment and a negative segment, respectively.

Discard the cycles and replace each path by an edge between the segments containing its endpoints, assigning the distance between those segments as its weight. This distance is at most the distance between the endpoints, which is at most the cost of the path by the triangle inequality. Thus the resulting bipartite multigraph has total edge weight at most $B+A_i$. Since each nonzero segment contains exactly $h$ unmatched points, its corresponding vertex has degree exactly $h$.

This $h$-regular bipartite multigraph decomposes into $h$ perfect matchings. To see this, take any set of positive-segment vertices. The number of edges incident to them is $h$ times their number, while each neighboring negative-segment vertex is incident to at most $h$ of these edges. Therefore, the number of neighboring negative-segment vertices is at least the number of chosen positive-segment vertices. By Hall's theorem, a perfect matching exists. Removing it decreases every vertex degree by one, and repeating this argument gives the required decomposition. At least one of these $h$ perfect matchings has cost at most $(B+A_i)/h$. The minimum-weight perfect matching between segments costs no more, so
\[
M_i\leq\frac{B+A_i}{h}.\qedhere
\]
\end{proof}

\paragraph{Averaging over partitions.}
We now bound the total cost of the auxiliary matchings constructed in the preceding lemma. Their cost is controlled by the alternating sequences associated with the stack positions: in a nonzero segment, the matching cost for each sequence is at most half the length of the subpath within the segment from its first point to its last point; in a zero segment, it is at most the full length of that subpath. We charge these costs to the edges of the reference tour and sum over the partitions.

\begin{lemma}[Total cost of the auxiliary matchings]\label{lem:auxiliary-matching-total-cost}
Under the hypothesis of Lemma~\ref{lem:matching-cost}, let $A_i$ be the cost of the auxiliary matching constructed in Lemma~\ref{lem:fixed-partition-matching} for the $i$th partition. Then
\[
\sum_{i=0}^{h-1}A_i\leq\frac{h(h-1)}2T.
\]
\end{lemma}

\begin{proof}
If $h=1$, each segment consists of a single service point, so the auxiliary matchings are empty and the claim follows. Henceforth assume $h\geq2$.

We first relate the alternating sequences to the global cumulative net supply $F$. Assign each service point an integer label by selecting one of the $F$-values on its preceding and following edges, according to the rule below. Segment types and the signs of the proper prefixes of zero segments are determined in the fixed direction of the reference tour.
\begin{center}
\begin{tabular}{@{}ll@{}}
\hline
Segment type & Selected $F$-value\\
\hline
Positive & Smaller value\\
Negative & Larger value\\
Zero with positive nonempty proper prefixes & Larger value\\
Zero with negative nonempty proper prefixes & Smaller value\\
\hline
\end{tabular}
\end{center}
Pushing to and popping from a fixed stack position correspond to crossings of the same unit interval by the cumulative net supply. Thus all points in one alternating sequence have the same integer label, and different stack positions have different labels. When a zero segment is processed in reverse, labels are still determined in the fixed direction of the reference tour. Reversal only reverses the order of the sequence; it does not change its point set or the subpath between its first and last points.

If the cut edge preceding a segment has $F$-value $c$, then the labels in a positive segment are $c,c+1,\ldots,c+h-1$, and those in a negative segment are $c,c-1,\ldots,c-h+1$. In a zero segment, because the opposite endpoint of each unit interval is selected, the labels lie strictly between $c$ and $c+h$, or strictly between $c-h$ and $c$. Consequently, each segment has at most one sequence whose label belongs to any given residue class modulo $h$. In the $i$th partition, the sequence with label congruent to $i$ in a nonzero segment consists only of its first point, since the net supply of each nonempty proper prefix lies strictly between the two cut values. A zero segment has no such sequence. Hence sequences with labels congruent to $i$ contribute no subpath cost.

Now fix an edge $e$ of the reference tour, and let $F(e)\equiv\rho\pmod h$. The $\rho$th partition cuts $e$, so no subpath within one of its segments charges this edge. For each $i\ne\rho$, let $S_i(e)$ be the segment containing $e$, bounded by the nearest cut edges of class $i$ on either side of $e$. To compare these segments by inclusion, view them as intervals along the tour unrolled around $e$.  Suppose that $x$ of these $h-1$ segments are zero and the remaining $h-1-x$ are nonzero.

Each $S_i(e)$ has at most one sequence whose label belongs to class $\rho$. Even if all the corresponding subpaths contain $e$, their total charge coefficient on $e$ is at most
\[
\frac{h-1-x}{2}+x.
\]

Next consider a sequence in $S_i(e)$ whose label belongs to a class $j\ne\rho$ and whose subpath contains $e$. We also have $j\ne i$, since a sequence of class $i$ has a subpath of length zero. The sequence contains service points on both sides of $e$, each incident to an edge of class $j$. Since $j\notin\{i,\rho\}$, these class-$j$ edges can be neither the boundary edges of class $i$ nor $e$. Thus the nearest edges of class $j$ on both sides of $e$ lie strictly inside $S_i(e)$, and
\[
S_j(e)\subsetneq S_i(e).
\]
Moreover, the contained segment $S_j(e)$ must be a zero one. Otherwise, its boundary $F$-values differ by $h$ or $-h$, and the unit changes of $F$ must pass through a value of class $i$ in its interior. This would give another cut edge of class $i$ inside $S_i(e)$, contradicting its definition.

Therefore, apart from sequences of class $\rho$, each charge corresponds to a zero segment strictly contained in another segment. If the containing segment is nonzero, there are at most $x(h-1-x)$ such ordered pairs, each contributing a charge coefficient of $1/2$. If the containing segment is also zero, two distinct segments cannot strictly contain each other, so there are at most $x(x-1)/2$ such ordered pairs, each contributing a charge coefficient of $1$. Each pair corresponds to at most one sequence, since a segment has at most one sequence in any given residue class.

Combining these charges, the total charge coefficient on $e$ over all partitions is at most
\[
\begin{aligned}
\frac{h-1-x}{2}+x
+\frac{x(h-1-x)}2
+\frac{x(x-1)}2
&=\frac{(h-1)(x+1)}2\\
&\leq\frac{h(h-1)}2,
\end{aligned}
\]
where the last inequality uses $0\leq x\leq h-1$. Multiplying this coefficient by the edge cost $c(e)$ and summing over all edges of the reference tour gives
\[
\sum_{i=0}^{h-1}A_i\leq\frac{h(h-1)}2T.\qedhere
\]
\end{proof}

\begin{proof}[Proof of Lemma~\ref{lem:matching-cost}]
Under the hypothesis of Lemma~\ref{lem:matching-cost}, all $h$ partitions exist and contain nonzero segments. By Lemmas~\ref{lem:fixed-partition-matching} and~\ref{lem:auxiliary-matching-total-cost},
\[
\min_{0\leq i<h}M_i
\leq\frac1h\sum_{i=0}^{h-1}M_i
\leq\frac Bh+\frac1{h^2}\sum_{i=0}^{h-1}A_i
\leq\frac Bh+\frac{h-1}{2h}T.
\]
Thus at least one partition satisfies the required matching cost bound.
\end{proof}

\subsection{Lower Bounds for Preemptive Tours}\label{app:lower-bounds}

This subsection proves the two lower bounds used in Sections~\ref{sec:analysis}. Let $T^*$ be the cost of an optimal TSP tour through all positive and negative points, and let $B$ be the cost of a minimum-weight perfect matching between the original positive and negative points.

\begin{lemma}[Lower bounds for preemptive tours]\label{lem:preemptive-lower-bounds}
For any feasible preemptive tour of cost $C$,
\[
T^*\leq C,\qquad 2B\leq kC.
\]
\end{lemma}

\begin{proof}
Select one occurrence of each positive and negative point on the given tour, and shortcut the tour to visit only these selected occurrences in cyclic order. By the triangle inequality, the resulting TSP tour costs at most $C$, so $T^*\leq C$.

To prove the second inequality, subdivide the route at every pickup, delivery, and temporary drop or retrieval, so that the vehicle's load remains constant during each travel step. Represent each travel step by a directed edge oriented in the direction of travel and having cost equal to its length. Different visits to the same location correspond to the same vertex, and repeated traversals are retained as parallel edges.

If the vehicle carries $q$ items during a travel step, add $q$ copies of the corresponding directed edge to a first directed multigraph. Each positive point supplies one item, each negative point receives one item, and all items temporarily stored at intermediate locations are eventually retrieved. Consequently, in this graph, each positive point has outdegree one greater than its indegree, each negative point has indegree one greater than its outdegree, and every other location has equal indegree and outdegree. Even when temporary storage occurs at a positive or negative point, temporary drops and subsequent retrievals cancel and do not affect this difference.

Construct a second directed multigraph by adding $k-q$ reversed copies of the edge for the same travel step. Since the original route is closed, the number of arrivals at each location equals the number of departures, so assigning $k$ forward copies to each travel step yields zero net outflow at every location. Thus, before reversing the edges, the outdegree minus indegree contributed by the $k-q$ copies of remaining capacity is the negative of that contributed by the $q$ copies of actual load. Reversing all edges gives the second graph the same degree differences as the first. Hence it also has one unit of net outflow at each positive point, one unit of net inflow at each negative point, and balanced inflow and outflow at every other location.

For either graph, first delete directed cycles. Then trace the remaining directed edges from a positive point that still has positive net outflow until reaching a negative point that still has positive net inflow, obtaining a directed path between them. Deleting this path leaves the degree differences at its internal vertices unchanged and balances its two endpoints. Repeating this process decomposes the remaining edges into paths from positive to negative points, with each positive point starting exactly one path and each negative point ending exactly one path. Replacing each path by an edge between its endpoints gives a perfect matching between the original positive and negative points. By nonnegativity of edge costs and the triangle inequality, this matching costs no more than the total edge cost of the graph. Therefore, each of the two graphs has total cost at least $B$.

For each travel step of the original route, the two graphs together contain $q+(k-q)=k$ copies of its edge. Since distances are symmetric, reversing an edge does not change its cost. Thus the sum of the costs of the two graphs is exactly $kC$, and
\[
2B\leq kC.\qedhere
\]
\end{proof}

\clearpage
\section{Implementation Details}\label{app:implementation}

The time and space bounds in Section~\ref{sec:analysis} assume that distances between service points are given by a distance matrix. Time is measured by the number of basic arithmetic operations and comparisons, and space by the number of stored scalar values. We provide further details on enumerating distinct partitions, outputting the route, handling rational distances, and converting a given preemptive tour.

For $k\geq2$, let $h=\lfloor k/2\rfloor$. By Appendix~\ref{app:matching-proof}, after fixing a traversal direction of the reference tour, the distinct partitions correspond one to one to the residue classes modulo $h$ that occur among the global cumulative net supplies $F(e)$. During a scan of the tour, record one representative edge for each class that occurs, and use the service point immediately after that edge as the reference point for its partition. This preprocessing takes $O(N)$ time. We then process only the partitions associated with these representatives, avoiding repeated matching computations for the same partition.

During the recursive assembly, record the current position in the service order of each active cycle. Suspend its service when visiting a child cycle along the spanning tree, and resume from that position upon return. Append each travel or service operation directly to the output sequence when it is performed, without repeatedly copying previously constructed routes for subtrees. Each nonzero segment is traversed once for complete service and at most once more along a subpath without service; each matching edge is traversed once, and each gap at most twice. Thus, the operation sequence for each candidate tour has length $O(N)$, and both generating the sequence and computing its cost take $O(N)$ time.

For nonnegative rational distances, let $\ell$ be the total binary encoding length of the distance matrix. Multiplying all distances by the product of their denominators yields integer weights of $O(\ell)$ bits each. This common scaling preserves comparison results, minimum-weight matchings, and approximation ratios. Minimum-weight perfect matchings with integer weights can be computed by exact polynomial-time algorithms~\citep{Gabow90}, while all costs used in the other steps are sums of polynomially many integer weights. Hence, the algorithm also has polynomial time and space complexity in the binary computation model.

For the conversion in Section~\ref{sec:preemption-gap}, let $L$ be the number of travel and service operations in the supplied preemptive tour. Scanning the tour in order and recording the first occurrence of each service point extracts a reference TSP tour in $O(L)$ time. We then apply the same partitioning, matching, and assembly procedure. Thus, under the operation-counting model above, the total conversion time is $O(L+N^3)$, which is polynomial in the problem instance and the representation length of the supplied tour.

\end{document}